\documentclass{basi}
\def\mc{meridional circulation}
\def\wald{Waldmeier effect}
\def\bl{Babcock--Leighton}
\def\mm{Maunder minimum}

\usepackage[british]{babel}
\usepackage[varg]{txfonts}

\def\lsim{\lower.5ex\hbox{$\; \buildrel < \over \sim \;$}}
\def\gsim{\lower.5ex\hbox{$\; \buildrel > \over \sim \;$}}
\def\ch{\lower-0.55ex\hbox{--}\kern-0.55em{\lower0.15ex\hbox{$h$}}}
\def\lh{\lower-0.55ex\hbox{--}\kern-0.55em{\lower0.15ex\hbox{$\lambda$}}}

\begin{document}

\title[Modeling irregular solar cycle]{Modeling the irregularities of solar cycle using flux transport dynamo models}
\author[B.~B.\ Karak]{Bidya Binay Karak\thanks{Email: \texttt{bidya$\_$karak@physics.iisc.ernet.in}}\\
Department of Physics, Indian Institute of Science, Bangalore 560012, India}

\date{Received --- ; accepted ---}

\maketitle

\label{firstpage}
\begin{abstract}
The sunspot number varies roughly periodically with time. However the
individual cycle durations and the amplitudes are found to vary in an
irregular manner. It is observed that the stronger cycles are having shorter
rise times and vice versa. This leads to an important effect know as the
Waldmeier effect. Another important feature of the solar cycle
irregularity are the grand minima during which the activity level is
strongly reduced. We explore whether these solar cycle irregularities can be
studied with the help of the flux transport dynamo model of the solar
cycle. We show that with a suitable stochastic fluctuations in a 
regular dynamo model, we are able to reproduce many irregular features
of the solar cycle including the Waldmeier effect and the grand minimum. 
However, we get all these results only if the value of the turbulent 
diffusivity in the convection zone is reasonably high.

\end{abstract}

\section{INTRODUCTION}
Although the sunspot number varies periodically with time with an average
period of 11 year, the individual cycle period (length) and also the strength (amplitude) vary 
in a random way. It is observed that the stronger cycles have shorter periods
and vice versa. This leads to an important feature of solar cycle known as 
Waldmeier effect. It says that there is an anti-correlation between the rise 
time and the peak sunspot number. We call this as WE1. Now instead of 
rise time if we consider the rise rate then we get very tight positive correlation 
between the rise rate and the peak sunspot number. We call this as WE2.

Another important aspect of solar activity are the grand minima. These are 
the periods of strongly reduced activity. A best example of these is the 
\mm\ during during 1645--1715. It was not an artifact of few observations, 
but a real phenomenon (Hoyt \& Schatten 1996). From the study of the 
cosmogenic isotope $^{14}$C data in tree rings, Usoskin et al. (2007) 
reported that there are $27$ grand minimum during last $11,000$ years. 

\section{METHODOLOGY AND RESULTS}
We want to model these irregularities of solar cycle using flux transport dynamo
model (Choudhuri et al. 1995; Dikpati \& Charbonneau 1999; Chatterjee et al. 2004). 
In this model, the turbulent diffusivity 
is an important ingredient which is not properly constrained. 
Therefore several groups use different value of diffusivity and 
this leads to two kinds of flux transport dynamo model -- high diffusivity 
model and low diffusivity model. In the earlier model, the value of 
diffusivity usually used is $\sim 10^{12}-10^{13}$ cm$^2$ s$^{-1}$ (see also
Jiang et al. 2007 and Yeates et al. 2008 for details), whereas 
in the latter model, it is $\sim 10^{10}-10^{11}$ cm$^2$ s$^{-1}$. 
We mention that the mixing length theory gives the 
value of diffusivity as $\sim 10^{12}$ cm$^2$ s$^{-1}$. 
Another important flux transport agent in this model is 
the meridional circulation. Only since 1990's we have some observational 
data of meridional circulation near the surface and therefore we do not 
know whether the \mc\ varied largely with solar cycle in past or not. 
However if the flux transport dynamo 
is the correct dynamo for the solar cycle, then one can consider the solar 
cycle period variation as the variation for the \mc\ because the 
cycle period is strongly determined by the strength of the meridional circulation in this model.
Now the periods of the solar cycle indeed had much variation in past, 
then we can easily say 
that the \mc\ had significant variation with the solar cycle. Therefore 
the main sources of randomness in the flux transport dynamo model are the stochastic fluctuations in 
\bl\ process of generating poloidal field and the stochastic fluctuations in the meridional circulation.
In this paper we explore the effects of fluctuations of the latter.

\subsection{Modeling last $23$ solar cycles}
\begin{figure}
\centering
\includegraphics[width=1.00\textwidth]{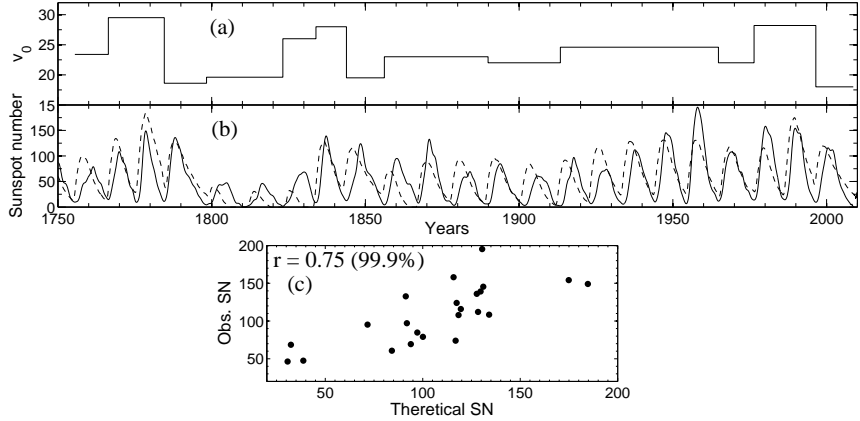}
\caption{(a) Variation of amplitudes of \mc\ $v_0$ (in m~s$^{-1}$) with time 
(in yr). The solid line is the 
variation of $v_0$ used to match the theoretical periods with the 
observed periods. (b) Variation of theoretical sunspot number (dashed 
line) and observed sunspot number (solid line) with time. (c) Scatter diagram showing 
peak theoretical sunspot number and peak observed 
sunspot number. The linear correlation coefficients and the corresponding 
significance levels are given on the plot.}
\label{fit23}
\end{figure}

We model last $23$ cycles by fitting the periods with variable 
meridional circulation in a high diffusivity model based on Chatterjee et al. (2004) model. 
The solid line in Fig.~\ref{fit23}(a) shows the variation of the amplitude of \mc\ $v_0$
used to model the periods of the cycles.
Note that we did not try to match the periods of each 
cycles accurately which is bit difficult. We change $v_0$ between two cycles 
and not during a cycle. In addition, we do not change $v_0$ if the 
period difference between two successive cycles is less than 
$5\%$ of the average period.

In Fig.~\ref{fit23}(b), we show the theoretical sunspot series (eruptions) by dashed line along
with the observed sunspot series by solid line. The theoretical
sunspot series has been multiplied
by a factor to match the observed value. It is very interesting to see that most of the 
amplitudes of the theoretical sunspot cycle have been matched with the observed sunspot cycle.
Therefore, we have found a significant correlation between these two (see Fig.~\ref{fit23}(c)). 
This study suggests that a major part of the fluctuations of the amplitude of the solar cycle
may come from the fluctuations of the meridional circulation. This is a very important 
result of this analysis.

Now we explain the physics of this result based on Yeates et al. (2008).
Toroidal field in the flux transport model, is generated by 
the stretching of the poloidal field in the tachocline. The
production of this toroidal field is more if the poloidal field remains
in the tachocline for longer time and vice versa. However, the poloidal field
diffuses during its transport through the convection zone. As a result, if
the diffusivity is very high, then much of the poloidal field diffuses away
and very less amount of it reaches the tachocline to induct toroidal field.
Therefore, when we decrease $v_0$ in high diffusivity model to match
the period of a longer cycle, the poloidal field gets more time to diffuse
during its transport through the convection zone. This ultimately leads
to a lesser generation of toroidal field and hence the cycle becomes
weaker. On the other hand, when we increase the value of $v_0$ to match 
the period of a shorter cycle, the poloidal field does not get much 
time to diffuse in the convection zone. Hence it produces stronger toroidal field and the cycle becomes stronger.
Consequently, we get weaker
amplitudes for longer periods and vice versa. However, this is not the case 
in low diffusivity model because in this model the diffusive decay of the fields are not
much important. As a result, the slower meridional circulation means that
the poloidal field remains in the tachocline for longer time and therefore
it produces more toroidal field, giving rise to a strong cycle. Therefore, we do not get
a correct correlation between the amplitudes of theoretical sunspot number and
that of observed sunspot number when repeat the same analysis in low diffusivity 
model based on Dikpati \& Charbonneau (1999) model.

\subsection{Modeling Waldmeier effect}
\begin{figure}[!h]
\centering
\includegraphics[width=1.0\textwidth]{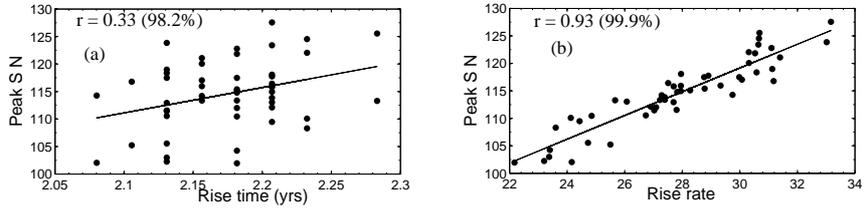}
\caption{Theoretical plots of WE1 (a) and WE2 (b) obtained
by introducing fluctuations in the poloidal field at the minima.}
\label{pol}
\end{figure}

We study the \wald\ using flux transport dynamo model. We have seen that 
the stochastic fluctuations in the \bl\ process and the stochastic fluctuations 
in the \mc\ are the two main sources of irregularities in this model. 
Therefore, to study \wald\ we first introduce suitable stochastic fluctuations 
in the poloidal field source term of \bl\ process. We see that this study cannot reproduce WE1 
(Fig.~\ref{pol}(a)). However it reproduces WE2 (Fig.~\ref{pol}(b)).

\begin{figure}[!h]
\centering
\includegraphics[width=1.0\textwidth]{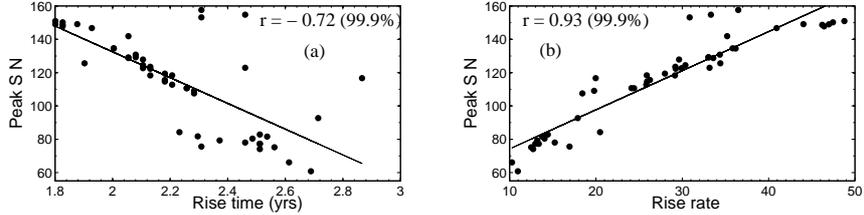}
\caption{Theoretical plots of WE1 and WE2 obtained by introducing fluctuations
in the meridional circulation.}
\label{mc}
\end{figure}

\begin{figure}[!h]
\centering
\includegraphics[width=1.0\textwidth]{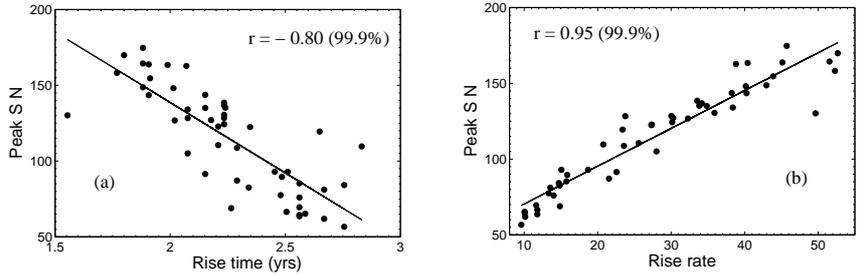}
\caption{Theoretical plots of WE1 and WE2 obtained by introducing fluctuations
in both the poloidal field generation and the meridional circulation.}
\label{both}
\end{figure}

Next we introduce stochastic fluctuations in the meridional circulation. 
Fig.~\ref{mc} shows this result. Interestingly, we see that it reproduces 
not only WE2, but also WE1 (see Fig.~\ref{mc}(a)).

Finally we introduce stochastic fluctuations in both the poloidal field source term 
and the meridional circulation. We see that both WE1 and WE2 are remarkably 
reproduced in this case (see Fig.~\ref{both}). We repeat the same study in low diffusivity model based on
Dikpati \& Charbonneau (1999) model. However in this case we are failed to reproduce 
WE1, only WE2 is reproduced. The details of this work can be found in 
Karak \& Choudhuri (2011).

\subsection{Modeling Maunder-like grand minimum}
We have realized that the \mc\ is important in modeling many aspects of 
solar cycle. Therefore we check whether a large decrease of the 
\mc\ leads to a Maunder-like grand minimum. To answer this question, we decrease $v_0$ to 
a very low value in both the hemispheres. We have done this in the 
decaying phase of the last sunspot cycle before Maunder minimum. 
We keep $v_0$ at low value for around 1~yr and then we again 
increase it to the usual value but at different rates in two 
hemispheres. In northern hemisphere, $v_0$ is increased at slightly 
lower rate than southern hemisphere.

\begin{figure}[!h]
\centering
\includegraphics[width=1.0\textwidth]{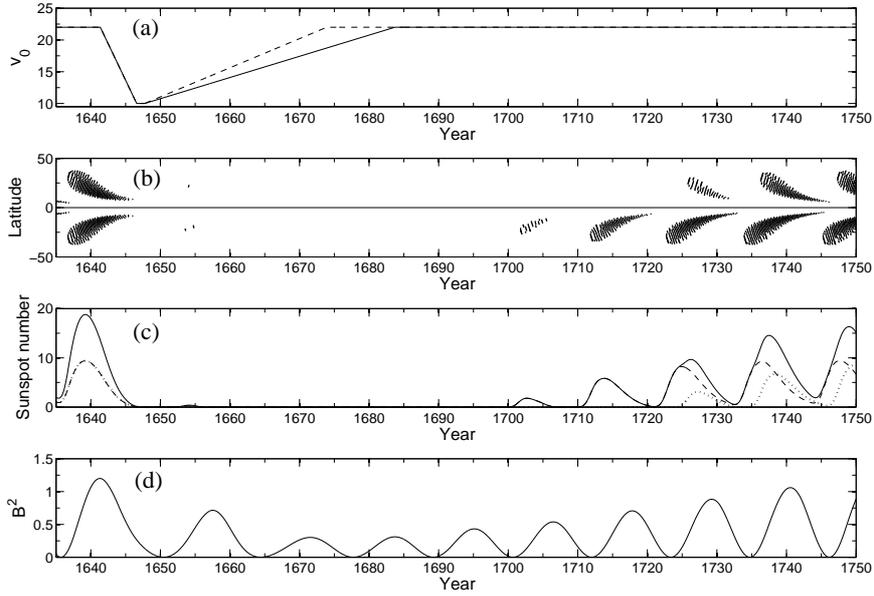}
\caption{(a) The solid and dashed line
show the variations of $v_0$ (in m~s$^{-1}$) in northern and
southern hemispheres with time. (b) The butterfly diagram. (c) The dashed
and dotted lines show the sunspot numbers in southern and northern hemispheres,
whereas the solid line is the total sunspot number. (d) Variation of energy density of toroidal field
at latitude 15$^{\circ}$ at the bottom of the convection zone.}
\label{mm}
\end{figure}

In Fig.~\ref{mm}, we show the theoretical results covering the Maunder
minimum episode.
Fig.~\ref{mm}(a), shows the maximum amplitude of meridional
circulation $v_0$ varied over this period in two hemispheres.
In Fig.~\ref{mm}(b), we
show the butterfly diagram of sunspot numbers, whereas
in Fig.~\ref{mm}(c), we show the variation of total sunspot number
along with the individual sunspot numbers in two hemispheres
(see the caption). In order to facilitate comparison
with observational data, we have taken the beginning of the year to be 1635.
Note that our theoretical results reproduce the sudden initiation and the 
gradual recovery, the North-South asymmetry of sunspot number observed in the
last phase of Maunder minimum and the cyclic oscillation of solar 
cycle found in cosmogenic isotope data.

We also mention that if we reduce the poloidal field to a very low 
value at the beginning of the Maunder minimum then also we can reproduce Maunder-like grand 
minimum (Choudhuri \& Karak 2009). However in both the cases, either we need to reduce the \mc\ or the poloidal 
field at the beginning of the Maunder minimum. However if we reduce the poloidal field
little bit, then one can reproduce Maunder-like grand minimum at a moderate value of
meridional circulation. The details of this study can be found in Karak (2010). 

We have shown that with a suitable stochastic fluctuations in 
the meridional circulation, we are able to reproduce many important 
irregular features of solar cycle including Waldmeier effect and 
Maunder like grand minimum. However we are failed to reproduce these
results in low diffusivity model. 
Therefore this study along with some earlier 
studies (Chatterjee, Nandy \& Choudhuri 2004; Chatterjee \& Choudhuri 2006; 
Goel \& Choudhuri 2009; Jiang, Chatterjee \& Choudhuri 2007; Karak 2010; 
Karak \& Choudhuri 2011; Karak \& Choudhuri 2012) 
supports the high diffusivity model for solar cycle.

Acknowledgements: I thank Prof. Arnab Rai Choudhuri for stimulating discussion 
and suggestion. I also thank the conference organizers for giving me the opportunity
to present my work.


\begin{thebibliography}{}
\bibitem{chatterjee}Chatterjee, P., Nandy, D., \& Choudhuri, A. R. 2004, A\&A, 427, 1019
\bibitem{ccj}Choudhuri, A.\ R., Chatterjee, P., \& Jiang, J., 2007, Phys. Rev. Lett., 98, 1103
\bibitem{karak}Choudhuri, A. R., \& Karak, B. B. 2009, RAA 9, 953
\bibitem{chou95}Choudhuri, A. R., Sch\"ussler, M., \& Dikpati, M. 1995, A\&A, 303, L29
\bibitem{dikpati99}Dikpati, M., \& Charbonneau, P. 1999, ApJ, 518, 508
\bibitem{jiang}Jiang, J., Chatterjee, P., \& Choudhuri, A. R. 2007, MNRAS, 381, 1527
\bibitem{hoyt}Hoyt, D. V., \& Schatten, K. H., 1996, Sol. Phys., 165, 181
\bibitem{karak10} Karak, B. B. 2010, ApJ, 724, 1021
\bibitem{karaknew} Karak, B. B., \& Choudhuri, A. R. 2011, MNRAS, 410, 1503
\bibitem{karak11}Karak, B. B., \& Choudhuri, A. R. 2012, Sol. Phys., 278:137
\bibitem{usos07}Usoskin, I. G., Solanki, S. K., \& Kovaltsov, G. A. 2007, A\&A, 471, 301
\bibitem{yeates}Yeates, A. R., Nandy, D., \& Mackay, D. H. 2008, ApJ, 673, 544
\end{thebibliography}
\end{document}